\documentclass[%
 reprint,
showpacs,preprintnumbers,
 amsmath,amssymb,
 aps,
]{revtex4-1}

\usepackage{graphicx}
\usepackage{dcolumn}
\usepackage{bm}

\def\al{\alpha}
\def\be{\beta}
\def\ga{\gamma}
\def\de{\delta}
\def\ep{\epsilon}

\def\la{\lambda}

\def\si{\sigma}

\def\Ga{\Gamma}

\def\La{\Lambda}

\def\mn{{\mu\nu}}

\def\cD{{\cal D}}
\def\cl{{\cal L}}

\def\frac#1#2{{\textstyle{{#1}\over {#2}}}}

\def\lsim{\mathrel{\rlap{\lower4pt\hbox{\hskip1pt$\sim$}}
    \raise1pt\hbox{$<$}}}
\def\gsim{\mathrel{\rlap{\lower4pt\hbox{\hskip1pt$\sim$}}
    \raise1pt\hbox{$>$}}}
\def\sqr#1#2{{\vcenter{\vbox{\hrule height.#2pt
         \hbox{\vrule width.#2pt height#1pt \kern#1pt
         \vrule width.#2pt}
         \hrule height.#2pt}}}}

\newcommand{\beq}{\begin{equation}}
\newcommand{\eeq}{\end{equation}}
\newcommand{\bea}{\begin{eqnarray}}
\newcommand{\eea}{\end{eqnarray}}

\begin{document}

\preprint{NCF/005}

\title{Vector Superfields and Lorentz Violation }

\author{Don Colladay}
 \altaffiliation[Also at ]{Division of Natural Sciences, New College of Florida.}
\author{Patrick McDonald}%
 \email{mcdonald@ncf.edu}
\affiliation{%
 New College of Florida\\
 5800 Bay Shore, Rd, \\
 Sarasota, FL 34243
}%

\date{\today}

\begin{abstract}
We extend Lorentz-violating Supersymmetry models to include vector superfields.  The 
CPT-preserving model generalizes easily, while the obvious attempt at 
generalizing the CPT-violating model meets serious obstructions.  
Generalizations of the CPT-preserving but Lorentz-Violating model to higher
dimensions are also straightforward.  Compactification is used to reduce the six-dimensional theory to an ${\cal{N}} =2$ Lorentz-violating theory in four dimensions, while the ten-dimensional 
theory is used to produce
an ${\cal{N}}=4$ Lorentz-violating theory.  This may be useful in future constructions involving
Ads/CFT correspondence with Lorentz violation.
\end{abstract}

\maketitle

\section{Introduction}
The conventional approach to applying supersymmetric theories to nature is to first
extend the Poincar\'e symmetry to a superalgebra, then break the supersymmetry
to reproduce something resembling the standard model at low energies.  
Alternatively, one may consider preserving the supersymmetry and breaking the Lorentz
symmetry of the theory.
Attempts have been made in this direction using two main approaches.  
The first involves perturbing the superalgebra sector involving $Q$ and $P^\mu$
operators, while the second involves keeping this sector of the superalgebra 
unperturbed.
In the second case, it is argued that it is not possible to include any operators into the 
minimal supersymmetric extension of the Standard Model  with dimension four or 
less\cite{posp}.
There is also a third approach in which the SUSY is explicitly broken along with the Lorentz symmetry by the background fields present in the supermultilplets\cite{belich}.

The first case in which the superalgebra is modified is explored in the current manuscript.
Previous work along these lines began with the chiral superfield model by Berger and
Kosteleck\'y \cite{bergkost} in which the derivative operator was twisted in various ways
to produce CPT-preserving and CPT-violating models of supersymmetry with Lorentz
Violation.
In the current manuscript, these models are extended to the vector superfield in
higher dimensions, and nonabelian theories.
The CPT-preserving model generalizes with minimal difficulty, while the CPT-violating
version of the theory meets serious impediments.  
Compactification of the higher-dimensional theories is used to produce four-dimensional
SUSY theories with ${\cal{N}}=2$ and ${\cal{N}}=4$ extended supersymmetries.
The ${\cal{N}} = 4$ theory is particularly interesting as it provides the first perturbed
version of the conformal field theory that may have a simple Ads/CFT dual.

\section{CPT preserving model}
The CPT-preserving model that violates Lorentz symmetry is particularly simple to 
implement as there exists a modified superspace implementation in terms of $Q$ operators.
This is not the case for the CPT-violating model.
Introduce the twisted derivative operator \cite{bergkost}
\beq
\tilde \partial_\mu = \partial_\mu + k_{\mu}^{~\nu} \partial_\nu ,
\eeq
the general vector superfield is expanded in terms of the twisted derivative as
\bea
V(x,\theta) & = & C(x) + i \overline \theta \gamma^5 \omega(x) - {i \over 2} \overline \theta \gamma^5
\theta M(x) - {1 \over 2} \overline \theta \theta N(x) \nonumber \\
& & + {1 \over 2} \overline \theta \gamma^5
\gamma^\mu \theta \tilde V_\mu 
- i \overline \theta \gamma^5 \theta \overline \theta \left[ 
\lambda(x) + {i \over 2}\tilde {\not \partial} \omega(x) \right] \nonumber \\
& & 
+ {1 \over 4} (\overline \theta \theta)^2\left( D(x) - {1 \over 2} \tilde \partial_\mu \tilde \partial^\mu C(x) \right) ,
\eea
where $\tilde V_\mu = V_\mu + k_{\mu\nu} V^\nu$ is inserted into the expansion.
This turns out to be necessary to preserve conventional gauge invariance in the 
resulting supersymmetric lagrangian.  It is also consistent with imposing a shift of the entire
covariant derivative operator rather than simply the partial derivative part.
The supersymmetry transformations are implemented using the operator
$\delta_Q V(x,\theta) = -i \overline \epsilon Q V(x,\theta)$, where
\beq
Q =  i \partial_{\overline \theta} - \gamma^\mu \theta \tilde \partial_\mu.
\eeq
Note that this operator satisfies the perturbed anticommuation relation
\beq
\{ Q, \overline Q \} =  2 \gamma^\mu (P_\mu + k_{\mu\nu}P^\nu).
\eeq
Matching terms in the conventional way yields the SUSY transformation properties of the fields.
The ones that will be useful in the present work are
\bea
\delta \lambda & = & {i \over 2} \si^\mn \tilde F_\mn \ep - i \gamma^5 D \epsilon, \\
\delta \tilde V^\mu & = & -i \overline \ep \left( \gamma^\mu \la + i \tilde \partial^\mu \omega \right), \\
\delta D & = & \overline \ep  \tilde {\not \partial} \gamma^5 \lambda ,
\eea
where the quantity $\tilde F$ is a twisted field strength (note that $\tilde F$ does
not have the conventional meaning of dual in this manuscript) given by
\beq
\tilde F^\mn = \tilde \partial^\mu \tilde V^\nu - \tilde \partial^\nu \tilde V ^\mu.
\eeq
Note that the $D$-term still transforms as a total derivative and can therefore be used
to construct SUSY lagrangians.

A general super gauge transformation is implemented by adding a chiral superfield expressed
in terms of the twisted derivative.
The result is the vector superfield in the Wess-Zumino gauge
\beq
V(x,\theta) = {1 \over 2} \overline \theta \gamma^5 \gamma^\mu \theta \tilde V_\mu 
- i \overline \theta \gamma^5 \theta \overline \theta \lambda(x) 
+ {1 \over 4} (\overline \theta \theta)^2 D(x)
\eeq
Note that a twisted gauge transformation on $\tilde V_\mu$ results in symmetry under the conventional gauge transformation for $V^\mu$
\beq
V^\mu \rightarrow V^\mu + \partial^\mu \Lambda.
\eeq
This is the reason that the original superfield was expanded in terms of $\tilde V^\mu$ rather
than $V^\mu$.
The supercovariant derivative can be used to calculate the spinor superfield
$W_\alpha = 1/4 \overline \cD \gamma^5 \cD \cD_\alpha V$, with 
$\cD = i  \partial_{\overline \theta} + \gamma^\mu \theta \tilde \partial_\mu$.
The left component of $W$ is
\bea
W_L & = & \lambda_L(x_+) - {1 \over 2}\gamma^\mu\gamma^\nu \theta_L \tilde F_{\mn}(x_+)
-i \overline \theta_R \theta_L \tilde {\not \partial} \lambda_R(x_+) \nonumber \\
& & 
- i \theta_L D(x_+) ,
\eea
where the left-chiral coordinate $x_+$ is given by
\beq
x^\mu_+ = x^\mu + {i \over 2}\overline \theta \gamma^5 \gamma^\mu \theta + 
{i \over 2}k^\mn \overline \theta \gamma^5 \gamma_\nu \theta .
\eeq
The lagrangian invariant under the modified SUSY transformation is constructed using
the $\theta^2$-term in the second-order function
\beq
\cl = {1 \over 2} {\rm Re}\left( \overline W_L \gamma^5 W_L |_{\theta^2} \right)= 
-{1 \over 4} \tilde F^2 + {i \over 2}\overline \la \tilde {\not \partial} \la + {1 \over 2}D^2.
\label{susyl}
\eeq
Note that this construction works because the left-superfield satisfies the twisted
chirality condition
\beq
\cD_R W_L = 0
\eeq
where $\cD_R = 1/2(1 + \gamma^5) \cD$ is the right-handed projection of the twisted supercovariant derivative operator. 
Any function of $W_L$ therefore satisfies the twisted chirality condition and hence leads to
a SUSY covariant $\theta^2$-term, or `$F$-Term' as it is commonly denoted.
Explicit verification of this SUSY requires the use of the twisted Jacobi Identity 
\beq
\ep^{\mn\al\be} \tilde \partial_\nu \tilde F_{\al\be} = 0 ,
\eeq
satisfied by the twisted field strength due to simple symmetry principles.
This Lagrangian can be put into the form of the Standard Model Extension (SME) Parameters
by expanding the Lagrangian in the form
\bea
\cl & = & -{1\over 4} (F^2 + k_F^{\mn\alpha\beta}F_{\mn}F_{\al\be}) + {i \over 2} (\overline \lambda
\ga^\mu \partial_\mu \lambda + \overline \lambda c_\mn \gamma^\mu \partial^\nu \lambda)
\nonumber \\
& &
+{1 \over 2}D^2,
\label{twistedsusy}
\eea
where 
\bea
k_F^{\mn\al\be} & = & 2 (2 k^{\al \mu}  + (k^2)^{\al\mu}) g^{\be\nu} + 4 (k^{\mu \al} + (k^2)^{\al\mu} )
k^{\nu\be} \nonumber \\
& & + (k^2)^{\al\mu} (k^2)^{\be\nu},
\eea
and 
\beq
c_\mn = k_\mn.
\eeq
In this expression, $k^2$ represents the square of the matrix for $k$
\beq
(k^2)^\mn = k^{\mu \al} k_{\al}^{~\nu}.
\eeq
Note that in contrast to the scalar chiral superfield model where only second-order
terms in $k$ were present, third- and fourth-order terms in $k$ appear in the vector
superfield Lagrangian.

\section{Attempt at construction of a CPT violating model}
CPT violation can be incorporated into the chiral superfield using a phase redefinition
of the fields involved \cite{bergkost}.  
When this same procedure is adopted for the vector superfield case, several problems
emerge.
Suppose that the fermion field is redefined by
\beq
\tilde \lambda = e^{-i \gamma^5 k \cdot x} \lambda
\eeq
to induce CPT-violating terms as is done in the chiral superfield case.
The phase cannot be absorbed into the real vector field $V_\mu$ as can be seen from
the following decomposition of the vector field supersymmetry transformation
\beq
\delta V^\mu = -i \overline \ep \gamma^\mu (e^{- i k \cdot x}  \lambda_R
+ e^{i k \cdot x} \lambda_L ) .
\eeq
This expression implies that it is impossible to absorb both of the phase changes into a 
single redefinition
of the vector field as was possible in the chiral superfield case.
This indicates that the CPT-violating generalization of the vector superfield probably
does not exist.

\section{${\cal{N}}=2$ FROM SIX-DIMENSIONAL THEORY}
The extension to higher dimensions and the following compactification process uses
the notation and techniques introduced in \cite{brink}.
The definitions of $L$ and $R$ are switched relative to the first half of this paper, 
this notation is kept in order to provide an easier comparison with the unperturbed results of
\cite{brink}.

The supersymmetric Lagrangian (\ref{susyl})
may be extended to six dimensions by allowing the $\gamma$ matrices and
vector field indices to run from 0 to 6 and setting the auxiliary field $D$ to zero.
It is not possible to impose the Majorana condition in six dimensions, but it is
possible to impose the Weyl condition.  The Lagrangian
\beq
\cl = -{1 \over 4} \tilde F^2 + i \overline \la \tilde {\not \partial} \la.
\eeq
is supersymmetric under the modified supersymmetry transformations
\bea
\delta \lambda & = & {i \over 2} \si^\mn \tilde F_\mn \ep , \\
\delta \tilde V^\mu & = & -i \left( \overline \ep \Ga^\mu \la  - \overline \la \Ga_\mu \ep \right) .
\eea
The gamma matrices are chosen in a basis where $\Gamma^\mu = \gamma^\mu \otimes I_2$
for $\mu=0,1,2,3$ and
\beq
\Gamma^{4,5} = \ga^5 \otimes i \si^{1,2},
\eeq
\beq
\Gamma^7 = \gamma^5 \otimes \si^3 .
\eeq
In this basis, the Weyl condition fixes $\la$ to have the structure
\beq
\la = \left(
\begin{array}{c}
L \chi \\
R \chi
\end{array}
\right),
\eeq
where $\chi$ is a four-dimensional Dirac spinor.
The components of the vector potential remain the same for $\mu=0,1,2,3$, while
$A_4 = S$ , $A_5 = P$, are relabeled as scalar and pseudoscalar fields.
It is convenient to investigate special cases for $k^\mn$.
The first case is when the lorentz violation is restricted to lie in the four physical
dimensions ($k^{\mu 4} = k^{\mu 5} = 0$).
The conventional dimensional reduction procedure yields the Lagrangian
\beq
\cl = -{1 \over 4} \tilde F^2 + {1 \over 2} \tilde \partial_\mu S \tilde \partial^\mu S
+ {1 \over 2} \tilde \partial_\mu P \tilde \partial^\mu P
+ i \overline \chi \not {\tilde \partial} \chi .
\eeq
This lagrangian has an ${\cal{N}}=2$ supersymmetry under the transformation laws
\bea
\delta \tilde V^\mu & = & i \left( \overline \alpha \ga_\mu \chi - \overline \chi \ga_\mu \al\right), \\
\de P & = & \overline \chi \ga^5 \al - \overline \alpha \ga^5 \chi, \\
\de S & = & i \left( \overline \chi \al - \overline \al \chi \right), \\
\de \chi & = & \left( {i \over 2} \si^\mn \tilde F_\mn \al + i \not{\tilde \partial} P \ga^5 
- \not{\tilde \partial} S \right) \al.
\eea
Another interesting special case is when the Lorentz violation is restricted to the
compactified dimensions.  In this case, the gauge and fermion terms are unperturbed
while the $S$ and $P$ fields are mixed by the $k^\mn$ parameters according to
\bea
\tilde S & = & (1 - k_{44})S - k_{45} P , \\
\tilde P & = & (1 - k_{55}) P - k_{45} S.
\eea
The lagrangian in this case is the usual one with the replacements
$S \rightarrow \tilde S$ and $P \rightarrow \tilde P$.
In the free theory this is just a redefinition of fields, but if the theory is coupled to other
sectors that involve the $SU(2)$ symmetry, there could be observable effects
of the violation.

\section{${\cal{N}}=4$ FROM TEN-DIMENSIONAL THEORY}

The supersymmetric Lagrangian (\ref{susyl})
 may be extended to ten dimensions by allowing the $\gamma$
matrices and vector field to have indices running from 0 to 9 and setting the auxiliary field
$D$ to zero
\beq
\cl = -{1 \over 4} \tilde F^2 + {i \over 2}\overline \la \tilde {\not \partial} \la.
\eeq
The gamma matrices are $32 \times 32$ matrices in this dimension and are 
denoted using $\Gamma^\mu$.
The first four ($\mu=0,1,2,3$) are chosen to reduce to the usual four-dimensional gamma matrices 
following compactification $\Gamma^\mu = \gamma^\mu \otimes I_8$.
The rest of the gamma matrices are defined as linear combinations of a more convenient
basis selected for its SU(4) properties
\bea
\Gamma^4 & = & \Ga^{14} + \Ga^{23} , \quad i \Gamma^7  =  \Ga^{14} - \Ga^{23} , \\
\Gamma^5 & = & \Ga^{24} - \Ga^{13}, \quad i \Gamma^8 = \Ga^{24} + \Ga^{13}, \\
\Gamma^6 & = & \Ga^{34} + \Ga^{12}, \quad i \Ga^9 = \Ga^{34} - \Ga^{12}.
\eea
where
\beq
\Ga^{ij} = \ga_5 \otimes 
\left( 
\begin{array}  {cc}
	0 & \rho^{ij} \\
	\rho_{ij} & 0 
\end{array}
\right),
\eeq
are written in terms of the $4 \times 4$ matrices
\beq
(\rho^{ij})_{kl} = \de_{ik} \de_{jl} - \de_{jk} \de_{il},
\eeq
and
\beq
(\rho_{ij})_{kl} = {1 \over 2} \ep_{ijmn}(\rho^{mn})_{kl} = \ep_{ijkl}.
\eeq
The product of all of the gamma matrices defines 
\beq
\Ga^{11} =  \gamma^5 \otimes \left( 
\begin{array}  {cc}
	I_4 & 0 \\
	0 & -I_4 
\end{array}
\right),
\eeq
and the charge conjugation matrix is
\beq
C^{10} =  C \otimes \left( 
\begin{array}  {cc}
	0 & I_4 \\
	I_4 & 0 
\end{array}
\right),
\eeq
where $C $ is the usual four-dimensional charge conjugation matrix.
The spinor is required to be an eigenstate of both $\Ga^{11}$ and $C_{10}$ 
therefore imposing the Majorana and Weyl conditions as in the standard case.

The previous construction of the perturbed theory still works, provided that the twisted Jacobi
Identity is modified appropriately to
\beq
A^{\nu\al\be} \tilde \partial_\nu \tilde F_{\al\be} = 0 ,
\eeq
where $A^{\nu \al \be}$ is a totally antisymmetric 3-tensor in 10 dimensions.
It is then possible to compactify the extra six dimensions onto a torus \cite{brink} and 
obtain a perturbed $N=4$ supersymmetric model that violates Lorentz Invariance through the twisted derivative.  Any Lorentz violation present in the extra dimensions will show up as a violation of the $SU(4)$ symmetry.  Such a model should be interesting from the perspective of Ads/CFT theory as it provides the first example of an appropriate Lorentz-violating theory that can be used in this context.  In fact, standard dimensional reduction procedures may be used to produce different
SUSY theories in four dimensions that may or may not violate Lorentz invariance in
the four physical dimensions.

For example, The $N=4$ SUSY model with Lorentz violation takes the form
\beq
\cl = -{1 \over 4} \tilde F^2 + i \overline \chi_i \tilde {\not \partial} L \chi^i 
+ {1 \over 4} \tilde \partial_\mu 
\phi_{ij} \tilde \partial^\mu \phi^{ij},
\eeq
where the $k_\mn$ background tensor was chosen to vanish along the compactified dimensions, but be of general form in the remaining four dimensions.
The complex scalar fields $\phi_{ij}$ transform as a {\bf 6} of $SU(4)$ and 
are defined according to 
\beq
\phi_{i4} = {1 \over \sqrt{2} }(A_{i+3} + i A_{i+6}), \quad i=1,2,3,
\eeq
and
\beq
\phi^{jk} = {1 \over 2} \epsilon^{jklm} \phi_{lm} = (\phi_{jk})^*,
\eeq
and the fermions $\chi^i$ form a {\bf 4}  representation of $SU(4)$.
The modified SUSY transformation laws of the fields are
\bea
\delta \tilde A_\mu & = & - i (\overline \alpha_i \gamma_\mu L \chi^i 
- \overline \chi_i \gamma_\mu L \alpha^i) \\
\delta \phi_{ij} & = & - \sqrt{2} i (\overline \alpha_j R \tilde \chi_i - \overline \alpha _i R \tilde \chi_j
 + \ep_{ijkl} \overline {\tilde \alpha}^k L \chi^l) \\
 \delta L \chi^i & = & {i \over 2} \si^\mn \tilde F_\mn L \alpha^i
 - \sqrt{2} \gamma^\mu \tilde \partial_\mu \phi^{ij} R \tilde \alpha_j \\
 \delta R\tilde  \chi_i & = & {i \over 2} \si^\mn \tilde F_\mn R\tilde \alpha_i
 + \sqrt{2} \gamma^\mu \tilde \partial_\mu \phi_{ij} L \alpha^j
\eea
Perturbations appear in all three sectors and they are connected by the 
requirement of supersymmetry.

Another interesting special case occurs when the Lorentz-violating coefficients 
are taken to vanish along the four physical dimensions, but are allowed to be
arbitrary in the compactified dimensions.  In this case, the four-dimensional theory
is Lorentz invariant, but violates $SU(4)$ symmetry.  The perturbations all appear in
the scalar sector as the resulting Lagrangian takes the form
\beq
\cl = -{1 \over 4} F^2 + i \overline \chi_i {\not \partial} L \chi^i 
+{1 \over 4} \partial_\mu 
\tilde \phi_{ij}  \partial^\mu \tilde \phi^{ij},
\eeq
where $\tilde \phi_{ij} = \phi_{ij} + \La_{ijkl} \phi_{kl}$ and 
the matrix $\La^{ijkl}$ contains the induced Lorentz violation from the higher dimensions.
The supersymmetry transformations for these fields are
\bea
\delta A_\mu & = & - i (\overline \alpha_i \gamma_\mu L \chi^i 
- \overline \chi_i \gamma_\mu L \alpha^i) \\
\delta \tilde \phi_{ij} & = & - \sqrt{2} i (\overline \alpha_j R \tilde \chi_i - \overline \alpha _i R \tilde \chi_j
 + \ep_{ijkl} \overline {\tilde \alpha}^k L \chi^l) \\
 \delta L \chi^i & = & {i \over 2} \si^\mn F_\mn L \alpha^i
 - \sqrt{2} \gamma^\mu \partial_\mu \tilde \phi^{ij} R\tilde  \alpha_j \\
  \delta R\tilde  \chi_i & = & {i \over 2} \si^\mn F_\mn R\tilde \alpha_i
 + \sqrt{2} \gamma^\mu \partial_\mu \tilde \phi_{ij} L \alpha^j.
\eea
In this case, identification of $\tilde \phi$ with the physical scalar fields removes
any effect of the higher-dimensional Lorentz violation and restores $SU(4)$ symmetry.  
However, this is only possible if the
fields $\phi_{ij}$ do not couple with other sectors as their redefinition would affect their symmetry
properties in relation to the other sectors.
If this is the case, the above model demonstrates a possible method by which the $SU(4)$
symmetry of an N=4 supersymmetry model may be broken.
Some fields can be driven to irrelevant status by making their kinetic coupling very small.
Note that the matrix $\La^{ijkl}$ in this model does not in fact have to be small since it 
does not induce Lorentz violation in the four physical dimensions.

\section{Non Abelian Generalization}

The above constructions also work if the full covariant derivative is twisted by $k^\mn$ 
according to
\beq
\tilde D^\mu = D^\mu + k^\mn D_\nu.
\eeq
For example, the non-Abelian ten-dimensional theory then takes the form
\beq
\cl = {\rm Tr} \left[ -{1 \over 2} \tilde F^2 
+ i \overline \la \tilde {\not D} \la \right] ,
\eeq
where $ i g \tilde F^\mn = [\tilde D^\mu, \tilde D^\nu]$,
and the supersymmetry transformations are the same as before written in
terms of $\tilde F$.
The key fact that allows the supersymmetry transformation to succeed is the
Bianchi Identity satisfied by $\tilde F^\mn$
\beq
[\tilde D^\al, [\tilde D^\be, \tilde D^\ga]] +  [\tilde D^\be, [\tilde D^\ga, \tilde D^\al]] +
[\tilde D^\ga, [\tilde D^\al, \tilde D^\be]] = 0 .
\eeq
The first case considered is when the Lorentz-violating couplings $k^\mn$ are restricted
to the four physical dimensions.  
The resulting Lagrangian is
\bea
\cl & = & {\rm Tr} \left[ -{1 \over 2} \tilde F^2 + 2i \overline \chi_i {\tilde {\not D}} L \chi^i 
\right.
\nonumber \\
& & \left.
+\sqrt{2} g \left( \overline{\tilde \chi}[ \phi_{ij}, L \chi^j] - 
\overline \chi [ \phi^{ij}, R \tilde \chi_j]
\right) \right.\nonumber \\
& & \left.
+{1 \over 2} \tilde D_\mu 
 \phi_{ij}  \tilde D^\mu \phi^{ij} + {g^2 \over 8} [\phi_{ij}, \phi_{kl}]^2
\right] .
\eea

The case when $k^\mn$ is restricted to the compactified dimensions is simplest,
the theory is the same as the conventional one with the replacement
\beq
\tilde \phi_{ij} = \phi_{ij} + \La_{ijkl} \phi_{kl},
\eeq
with lagrangian
\bea
\cl & = &{\rm Tr} \left[ -{1 \over 2} F^2 + 2i \overline \chi_i {\not D} L \chi^i \right. \nonumber \\
& &\left.
+\sqrt{2} g \left( \overline{\tilde \chi}[\tilde \phi_{ij}, L \chi^j] - 
\overline \chi [\tilde \phi^{ij}, R \tilde \chi_j]
\right) \right. \nonumber \\
& & \left.
+{1 \over 2} D_\mu 
\tilde \phi_{ij}  D^\mu \tilde \phi^{ij} + {g^2 \over 8} [\tilde \phi_{ij},\tilde \phi_{kl}]^2
\right].
\eea
\section{Note on Coordinate Transformations}
An interesting question arises as to the nature of the $k^\mn$ model regarding 
physical observability due to possible coordinate transformations that can alter, or 
remove Lorentz-Violating terms\cite{colmcjmp}.  
To answer this question, one starts with the conventional Lagrangian with no 
symmetry violating terms in it and performs the appropriate change in coordinates
for 
\beq
\partial^{\prime \mu} = \partial^\mu + k^\mu_{~ \nu}  \partial^\nu,
\label{derivtrans}
\eeq
corresponding to the derivative replacement used in the SUSY-LV formulas.
The initial lagrangian is written in the form
\beq
\cl_0 = - {1 \over 4}  g^{\mu\al} g^{\nu\be}F_\mn F_{\al\be} + {i \over 2}g^{\mn}\overline \la  
{\gamma_\mu \partial_\nu} \la + {1 \over 2} D^2
\label{convlag}
\eeq
where the metric $g^\mn = \eta^{\mn}$ is exhibited explicitly.
This form of the SUSY lagrangian is written in geometric form so that it is covariant under
the linear coordinate transformation consistent
 with Eq.\ (\ref{derivtrans}) ,
$x^\mu = x^{\prime \mu} - k^\mu_{~\nu} x^{\prime \nu}$.
A simple calculation to lowest-order in $k^\mn$ yields a modified metric
\beq
g^{\prime \mn} = \eta^\mn + 2 k^{\mn},
\eeq
and the corresponding lagrangian in the new coordiates reads the same as 
Eq.\ (\ref{convlag}) with primes on all of the quantities involved, including 
the gamma matrices which satisfy the modified Clifford algebra relation
\beq
\{ \ga^{\prime \mu}, \ga^{\prime \nu} \} = 2 g^{\prime \mn} = 2 \eta^\mn + 4 k^\mn.
\eeq
Expanding the $g^\prime$ factors and expressing the primed gamma in terms of the
conventional gamma yields the Lagrangian 
\bea
\cl & = & -{1\over 4} ( \eta^{\mu\al} \eta^{\nu\be}F^\prime_\mn F^\prime _{\al\be} 
+ k_F^{\mn\alpha\beta}F^\prime_{\mn}F^\prime_{\al\be}) \nonumber \\
& &
+ {i \over 2} (\overline \lambda^\prime
\eta^\mn \ga_\mu \partial^\prime_\nu \lambda^\prime + 
\overline \lambda^\prime k_\mn \gamma^{ \mu} \partial^{\prime \nu} \lambda^\prime)
+{1 \over 2}D^{\prime 2}.
\label{coordtranslag}
\eea
This lagrangian is formally equivalent to the supersymmetric Lorentz-breaking 
theory Eq.\ ( \ref{twistedsusy}), however, the vectors and spinors are resolved in a
non-orthogonal coordinate system
This means that all manipulations using the transformed lagrangian
must take into account this different property of the vector indices.
For example, quantities like
\beq
A^{\prime^\mu} A^\prime_\mu = g^{\prime\mn} A^\prime_\mu A^\prime_\nu
= \eta^\mn A^{\prime}_\mu A^\prime_\nu + 2k^\mn A^\prime_\mu A^\prime_\nu.
\eeq
must be used to compute lengths of vectors in the transformed system.  
Since the assumed metric in Eq.\ ( \ref{twistedsusy}) is simply $\eta^\mn$, no such
complication in dealing with vector components arises, so the theory is
physically distinct from a conventional theory expressed in skewed coordinates,
provided that the metric is accessible to the experimentalist through using rods and clocks
constructed of other, conventional Lorentz invariant fields.

Another way to put this is that it is possible to change coordinates to remove $k^\mn$
in the Lagrangian, but it will re-appear in the vector coordinate manipulations as
a pesky non-orthogonality of the vectors (and spinors) involved in the calculations of physical observables.
It will also appear in any Lorentz-invariant sectors as a new perturbation in those fields.
This argument also works provided there are at least two sectors with different valued $k_\mn$ tensors present in the theory.

\section{summary}

The main result of this paper is that the scalar CPT-preserving models with Lorentz Violation
easily generalize to the vector superfield and can be extended to the usual higher-dimensional
cases.  In addition, compactification can be used to construct extended supersymmetry models
in four-dimensions.  The theory may be abelian, or non-abelian, the same basic technique
works for both cases.
On the other hand, the scalar CPT-violating models do not easily generalize to the vector 
superfield.  The reason for this is connected with the inability to separate the vector potential
into left- and right-handed components.

The new ${\cal{N}}=4$ theories with Lorentz violation are particularly interesting from a 
theoretical viewpoint as the conventional theory plays a crucial role in Ads/CFT
theories.
One of the main purposes of this paper is to construct the perturbed ${\cal{N}}=4$ theories
so that they may be studied in this context.

\bibliography{susylvrev}

\end{document}